\documentclass[runningheads,a4paper]{llncs}

\usepackage{amssymb}
\usepackage{graphicx}
\usepackage{algorithm}
\usepackage{algorithmic}
\usepackage{hyperref}
\usepackage[numbers]{natbib}

\usepackage{amsmath}

\usepackage{url}
\urldef{\mailsa}\path|{Djallel.Bouneffouf}@it-sudparis.eu|

\begin{document}

\mainmatter  

\title{R-UCB: a Contextual Bandit Algorithm for Risk-Aware Recommender Systems}

\titlerunning{Lecture Notes in Computer Science. Authors' Instructions}

%
%
\author{Djallel Bouneffouf}
\authorrunning{Lecture Notes in Computer Science. Authors' Instructions}

\institute{Télécom SudParis, 9 Rue Charles Fourier, 91000 Evry, France\\
\mailsa \\
}

%
%

\toctitle{Lecture Notes in Computer Science}
\tocauthor{Authors' Instructions}
\maketitle

\begin{abstract}
Mobile Context-Aware Recommender Systems can be naturally modelled as an exploration/exploitation trade-off (exr/exp) problem, where the system has to choose between maximizing its expected rewards dealing with its current knowledge (exploitation) and learning more about the unknown user's preferences to improve its knowledge (exploration). This problem has been addressed 
by the reinforcement learning community but they do not consider the risk level of the current user's situation, where it may be dangerous to recommend items the user may not desire in her current situation if the risk level is high. 
We introduce in this paper an algorithm named R-UCB that 
considers the risk level of the user's situation
to adaptively balance between exr and exp. The detailed analysis of the experimental results reveals several important discoveries in the exr/exp behaviour.
\end{abstract}

\section{Introduction}
\label{intro}
User feedback (e.g., ratings and clicks) and 
situation ( 
e.g., location, time, near people) have 
become a crucial source of data for optimizing Mobile Context-Aware Recommender Systems (MCARS).
Knowledge about the environment must be accurately learned to avoid making undesired recommendations which may disturb the user in certain situations considered as critical or risky.

For this reason, the MCARS has to decide, for each new situation, whether so far learned knowledge should be exploited by selecting documents that appear more frequently in the corresponding user feedback, or if never seen documents should be selected in order to explore their impact on the user situation, increasing the knowledge about the environment.

Making exploration prevents from maximizing the short-term reward since exploration documents may yield to negative rewards, while exploitation of documents based on an uncertain environment knowledge can prevent from maximizing the long-term reward because document rating values may not be accurate.

This challenge is formulated as an exploration/exploitation (exr/exp) dilemma. 
One smart solution for exr/exp using  the "multi-armed bandit problem" is the hybrid approach done by \cite{21}. This approach combines the Upper Confident Bound (UCB) algorithm with the $\epsilon$-greedy algorithm. By introducing randomness into UCB, authors reduce the trouble in estimating confidence intervals.

This algorithm estimates both the mean reward of each document and the corresponding confidence interval. 
With the probability 1-$\epsilon$, this algorithm selects the document that achieves a highest upper confidence bound 
and, with the probability $\epsilon$, it uniformly chooses any other document. The $\epsilon$ parameter
essentially controls exr/exp.  
The problem is that it is difficult to decide in advance the optimal value of $\epsilon$.

We introduce in this paper an algorithm, named R-UCB, that computes the optimal value of $\epsilon$ by adaptively balancing exr/exp according to the risk of the user situation. 
We consider risky or critical a situation where it is dangerous to recommend uninteresting information for the user; this means that it is not desired, or even can yield to a trouble, that the user loses time reading a document which is not interesting for him in the current situation. 
In this case, the exploration-oriented learning should be avoided. 

R-UCB extends the UCB strategy with an update of exr/exp by selecting suitable user's situations for either exr or exp. We have tested R-UCB in an off-line evaluation with real data, as well as an on-line evaluation with professional mobile users.

The 
remaining of the paper is organized as follows. 
Section \ref{sec:related}
reviews related works. Section 
\ref{sec:crs} describes the algorithms involved in the proposed approach. The experimental evaluation is illustrated in Section 
\ref{sec:experimental}. The last section concludes the paper and points out possible directions for future work. 
\section{Related Work}
\label{sec:related}
We refer, in the following, a state of the art on MCARS and also techniques that tackle both making dynamic exr/exp (bandit algorithm) and considering the risk in the recommendation.

\subsection{MCARS}
\label{sec:22}
Few research works are dedicated to study MCARS. \cite{7} proposes a method which consists of building a dynamic user's profile based on time and user's experience. The user's preferences in the user's profile are weighted according to the situation (time, location) and the user's behavior. To model the evolution on the user's preferences according to his temporal situation in different periods (like workday or vacations), the weighted association for the concepts in the user's profile is established for every new experience of the user. The user's activity combined with the user's profile are used together to filter and recommend relevant content.

Another work \cite{13} describes a MCARS operating on three dimensions of context that complement each other to get highly targeted. First, the MCARS analyses information such as clients' address books to estimate the level of social affinity among the users. Second, it combines social affinity with the spatio-temporal dimensions and the user's history in order to improve the quality of the recommendations. 

In \cite{3}, the authors present a technique to perform user-based collaborative filtering. Each user's mobile device stores all explicit ratings made by its owner as well as ratings received from other users. Only users in  proximity are able to exchange ratings and they show how this provides a natural filtering based on social contexts.

In \cite{7} the authors propose a mobile recommender system for people in leisure time. The system predicts the current activity (eating,reading or shopping) from the context (time, location) and the behaviour of the user. The predicted activity combined with preference models of the user, are used to filter and recommend relevant content. To provide relevant advertisements to mobile users, authors in \cite{Bila} build a profile of each region visited by the user. Statistical techniques are then used to extract information from the visited regions, like "the frequency of visits", "duration and time of typical visits", and the user's profile is built on the basis of questionnaires.

Each work cited above tries to recommend interesting information to users taking account their contextual situation; however, they do not consider the exr/exp trade-off on their recommendations.

\subsection{Multi-armed Bandit Problem} 
\label{sec:multi}

Very frequently used in reinforcement learning to study 
exr/exp, the multi-armed bandit problem was originally described by \cite{16}. 
The $\epsilon$-greedy is one of the most used strategies to solve  
this problem. It chooses a random document with $\epsilon$-frequency, 
and chooses otherwise the document with the highest estimated mean, the estimation 
being based on the observed rewards. The $\epsilon$ is chosen by the user in the open interval ]0, 1[. 

The first variant of the $\epsilon$-greedy strategy is what \cite{6,10} refer to as the $\epsilon$-beginning strategy. This strategy makes exploration all at once at the beginning. For a given number I of iterations, documents are randomly selected during the $\epsilon$I first iterations; during the rest (1−-$\epsilon$)I iterations, the document of highest estimated mean is selected. Another variant of the $\epsilon$-greedy strategy is what \cite{10} calls the $\epsilon$-decreasing. In this strategy, the document with the highest estimated mean is always selected except when a random document is selected instead with $\epsilon_i$ frequency, where $\epsilon_i$ = {$\epsilon_0$/ i}, $\epsilon_0$ $\in$]0,1] and $i$ is the index of the current round. Besides $\epsilon$-decreasing, four other strategies was presented in \cite{3}. 

In contrast to the unguided exploration strategy adopted by $\epsilon$-greedy, another class of algorithms, known as UCB, use a smarter way to balance exr and exp. 
We can cite UCB methods in \cite{25} for rewards bounded in [0, 1] and the Price Of Knowledge Expected Reward (POKER) strategy \cite{26} for normally distributed rewards.

Both strategies construct a reward estimate for each document, which is the mean observed reward added to an additional coefficient of confidence intervals that is inversely related to the number of times  the document has been selected. 
The document with the highest estimated reward is selected. This way of reward estimation encourages exploration of documents that have been infrequently selected.
Another class of bandit algorithms based on Bayes rules (e.g., \cite{28})  
has good performance but they are computationally exorbitant \cite{29}. 

Few research works are dedicated to study the contextual bandit problem in recommender systems, 
considering the user's behaviour as the context. 

In \cite{13}, assuming the expected reward of a document is linear, they perform recommendation 
based on contextual information about the users' documents. To maximize the total number of user's clicks, this work proposes the LINUCB algorithm which is computationally efficient if the expected rewards of documents are linear which is not always the case.

In \cite{21}, the authors propose to solve bandit problem in dynamic environment by combining the UCB with the $\epsilon$-greedy strategy and they dynamically update the $\epsilon$ exploration value. At each iteration, they run a sampling procedure to select a new $\epsilon$ from a finite set of candidates. The probabilities associated to the candidates are uniformly initialized and updated with the Exponentiated Gradient (EG) \cite{7}. This updating rule increases the probability of a candidate $\epsilon$ if it leads to a user's click. Compared to both $\epsilon$-beginning and $\epsilon$-decreasing, this technique gives better results. 
\cite{13, 21} describe a smart way to balance exr/exp, but do not consider the user's situation and its associated risk during the recommendation.

We have already considered a part of these problems in \cite{BouneffoufBG12, bouneffouf2013drars, BouneffoufBG13} by modelling the MCARS as a contextual bandit algorithm. The proposed algorithm (contextual-$\epsilon$-greedy) classify the situations in two types: critical situations where the algorithm performs exploitation and non critical situations where the algorithm performs exploration. However, we have not considered the risk level of the situations in exr/exp trade-off.

\subsection{The Risk-aware Decision}
\label{sec:risk}

To the best of our knowledge, the risk-aware decision is not 
yet studied in recommender systems.
 
However, it has been studied for a long time in reinforcement learning, where  
the risk is defined as the reward criteria that  
takes into account not only the expected
reward, but also some additional statistics of the total reward, such as its variance or standard deviation \cite{31}.
The risk is measured with two types of uncertainties. The first, named parametric uncertainty, is related to the imperfect knowledge of the problem parameters. For instance, in the context of Markov decision processes (MDPs), \cite{32} propose to use the percentile performance criterion to control the risk sensitivity.

The second type, termed inherent uncertainty, is related to the stochastic nature of the system, like \cite{33}, who consider models where some states are error states  
representing a catastrophic result. 
More recently, \cite{34} developed a policy gradient algorithm for criteria that involves both the expected cost and the variance of the cost, and demonstrated the applicability of the algorithm in a portfolio planning problem. However, this work 
does not consider the risk of the situations in the exr/exp problem.

A recent work, \cite{30}, treated the risk 
and proposed the VDBE algorithm to extend $\epsilon$-greedy by introducing a state-dependent exploration probability, instead of hand-tuning a global parameter. The system  
makes exploration in situations when the knowledge
about the environment is uncertain, which is indicated by fluctuating action values during learning. In contrast, the amount of exploration is reduced as far as the system's knowledge becomes certain, which is indicated by very small or no value differences. 

We observe that the most common approach to define risk is through variance related criteria or the standard deviation adjusted 
reward; however, no work studied a more semantic definition of the risk nor studied the problem in MCARS.

\subsection{Main Contributions}
\label{sec:contribution}
As shown above, none of the mentioned works tackles
the exr/exp problem considering the semantic risk level of the situation.
This is precisely what we intend to do by exploiting the following new features: 

1) Handling semantic concepts to express situations and their associated risk level. The risk level is associated to a whole situation and/or the concepts composing the situation;  
 
2) Considering the risk level of the situation when managing 
exr/exp, which helps the MCARS adaptation to its dynamic environment. 
High exploration (resp. high exploitation) is achieved when the current user situation is "not risky" (resp. "risky"); 

3) Using off-line and on-line evaluations to measure the performances of the algorithm. 

We improve the extension of UCB with $\epsilon$-greedy (called here $\epsilon$-UCB) because it gives the best result in an off-line evaluation done by \cite{21}; however, our amelioration can be applied to any bandit algorithm. 

\section{The Proposed MCARS Model} 
\label{sec:crs}
This section focuses on the proposed model, starting by introducing the key notions used in this paper.

\textbf{\textit{Situation}:}  A situation is an external semantic interpretation of low-level context data, enabling a higher-level specification of human behaviour. More formally, a situation $S$ is a n-dimensional vector,  $S=(O_{\delta_{1}}.c_1,O_{\delta_{2}}.c_2,...,O_{\delta_{n}}.c_n)$ where each $c_i$ is a concept of an ontology $O_{\delta_{i}}$ representing a context data dimension. 
According to our need, we consider a situation as a 3-dimensional vector  $S=(O_{Location}.c_i, O_{Time}.c_j, O_{Social}.c_k)$ where $c_i, c_j, c_k $ are concepts of Location, Time and Social ontologies.
 
\textbf{ \textit{User preferences}:} User preferences $UP$ are deduced during the user navigation activities. $UP\subseteq D \times A \times V$ where $D$ is a set of documents, $A$ is a set of preference attributes and $V$ a set of values. We focus on the following preference attributes: \textit{click}, \textit{time} and \textit{recom} which respectively correspond to the number of clicks for a document, the time spent reading it and the number of times it was recommended.

\textbf{ \textit{The user model}:} The user model is structured as a case base composed of a set of situations with their corresponding $UP$, denoted $UM=\{(S^i; UP^i)\}$, where $S^i \in S$ is the user situation and $UP^i \in UP $ its corresponding user preferences.

We propose MCARS to be modelled as a contextual bandit problem including user's situation information.  
Formally, a bandit algorithm proceeds in discrete trials $t=1...T$. For each trial $t$, the algorithm performs the following tasks:

\textbf{Task 1:} Let $S^t$ be the current user's situation, and 
$PS$ the set of past situations. The system compares $S^t$ with the situations in \textit{PS} in order to choose the most similar one, $S^p$:
\begin{equation}
\label{eq:sp} 
S^p = argmax_{S^{i}\in PS}sim(S^t,S^i)
\end{equation} 

The semantic similarity metric is computed by:

\begin{equation}
 \label{eq:sim}
sim(S^t,S^i)=\frac{1}{|\Delta|}\sum_{\delta \in \Delta}sim_{\delta}(c^t_{\delta},c^i_{\delta})
\end{equation}                          
In Eq.~\ref{eq:sim}, $sim_{\delta}$ is the similarity metric related to dimension $\delta$ between two concepts $c_{\delta}^t$ and $c_{\delta}^i$, and $\Delta$ is the set of dimensions (in our case Location, Time and Social).
The similarity between two concepts of a dimension $\delta$  depends on how closely $c_{\delta}^t$ and $c_{\delta}^i$ are related in the corresponding ontology. To compute $sim_{\delta}$, we use the same similarity measure as \cite{15}:
\begin{equation}
\label{eq:simdelta}
sim_{\delta}(c_{\delta}^{t},c_{\delta}^{i})=2*\frac{depth(LCS)}{depth(c_{\delta}^t)+ depth(c_{\delta}^i)}
\end{equation}  
In Eq.~\ref{eq:simdelta}, $LCS$ is the Least Common Subsumer of $c_{\delta}^t$ and $c_{\delta}^i$, and $depth$ is the number of nodes in the path from the current node to the ontology root. 

\textbf{Task 2:} Let $D^p$ be the set of documents recommended in situation $S^p$. After retrieving $S^p$, the system observes rewards in previous trials for 
each document $d\in D^p$ in order to choose for recommendation the one with the greatest reward, which is the Click Through Rate (CTR) of a document. 
In Eq.~\ref{eq:frac}, the reward of document $d$, $r(d)$, is the ratio between the number of clicks ($v_i$) on $d$ and the number  of times $d$ is recommended ($v_j$). 

$\forall d \in D^p, UP^i$=$(d, click, v_i)$ $\in$ UP and $UP^j$=$(d, recom, v_j) \in$ UP we have: 
\begin{equation}
\label{eq:frac}
 r(d)=\frac{v_i}{v_j}
\end{equation} 
\textbf{Task 3: } 
The algorithm improves its  
document selection strategy with the new observation: in situation $S^t$, document $d$ obtains a reward $r(d)$. Depending on the similarity between the current situation $S^t$ and its most similar situation $S^p$, two scenarios are possible: 

(1) If sim($S^t$, $S^p$) $\neq 1$: the current situation does not exist in the case base; the system adds this new case composed of the current situation $S^t$ and the current user preferences $UP^t$; 

(2) If sim($S^t$, $S^p$) $= 1$: the situation exists in the case base; the system updates the case having premise the situation $S^p$ with the current user preferences $UP^t$.
\subsubsection{The $\epsilon$-UCB Algorithm.}
For a given situation, the algorithm recommends a predefined number of documents, specified by parameter \textit{N} using Eq.~\ref{alg:ucb}. 
Specifically, in trial $t$, this algorithm computes an index $b(d)= r(d)+c(d)$ for each document $d$, where: $r(d)$ (Eq. \ref{eq:frac}) is the mean reward obtained by $d$ and $c(d)$ is the corresponding confidence interval, so that $c(d)=\sqrt{\frac{2 \times log(t)}{v_j}}$ and $v_j$ is the number of times $d$ was recommended. 
With the probability 1-$\epsilon$, $\epsilon$-UCB selects the document with the highest upper confidence bound $d_t = argmax_{d \in D^p}b(d)$; and
with the probability $\epsilon$, it uniformly chooses any other document.
\begin{equation}
\label{alg:ucb}
d_t =\left\{ \begin{array}{rcl}
		 argmax_{d \in (D^p-RD)}b(d) & if \hspace{.1cm} q > \epsilon \\ 
              \hspace{-.5cm}  Random(D^p-RD)  & otherwise 
          \end{array}\right. 
\end{equation} 
In Eq.~\ref{alg:ucb}, $D^p$ is the set of documents included in the user's preferences $UP^p$ corresponding the most similar situation ($S^p$) to the current one ($S^t$); $RD$ is the set of documents to recommend; 
$Random()$ is the function returning a random element from a given set; \textit{q} is a random value uniformly distributed over [0, 1] which 
controls exr/exp; $\epsilon$ is the probability of recommending a random exploratory document. 

\subsubsection{The R-UCB Algorithm }
\label{subsub:rucb}
To improve the adaptation of the $\epsilon$-UCB algorithm (Alg.~\ref{alg:ucb}) to the risk level of the situations, the R-UCB algorithm computes the probability of exploration $\epsilon$, by using the situation risk level $R(S^t)$, as indicated in Eq.~\ref{eq:epsilon}. A strict exploitation ($\epsilon$=0) leads to a non optimal documents selection strategy, this is why $R$ is multiplied by $(1-\epsilon _{min})$, where $\epsilon_{min}$ is the minimum exploration allowed in $CS$ and $\epsilon_{max}$ is the maximum exploration allowed in all situations (these metrics are computed using a cross-validation).
\begin{equation}
 \label{eq:epsilon}
\epsilon = \epsilon_{max}-R(S^t)*(\epsilon_{max}-\epsilon_{min})
\end{equation} 
%
%

Note that we still consider an $\epsilon_{min}$ random exploration indispensable to avoid that document selection in $CS$ become less optimal. 

\begin{algorithm} [H]
   \caption{The R-UCB algorithm}
   \label{alg:RUCB} 
\begin{algorithmic}
   \STATE {\bfseries Input:} $S^t, D^t, D^p, RD = \emptyset, N, \epsilon _{min}, , \epsilon _{max}$
   \STATE  {\bfseries Output:} RD
   \STATE $\epsilon$ = $\epsilon_{max}-R(S^t)*(\epsilon_{max}-\epsilon_{min})$  //$R(S^t)$ is computed as described in Sec.~\ref{subsection:Risk}
     \STATE RD=$\epsilon$-UCB($\epsilon, D^p$, $D^t$, RD, N) 
   \\  
   \end{algorithmic}
\end{algorithm}
To summarize the algorithm R-UCB, the system makes a low exploration when the current user's situation is critical; otherwise, the system performs high exploration. In this case, the degree of exploration decreases when the risk level of the situation increases.

\subsection{Computing the Risk Level of the Situation}
\label{subsection:Risk}
In a contextual environment, the exploration-exploitation trade-off is directly related to the risk to upset the user (the risk level of the situation), this is why computing the risk level of the situation is indeed indispensable. 
 
As we observe from the state of the art, the best approach to compute the risk is from a hybrid approach that combines both the variance of the cost and the expected environment cost. However, hybrid approaches considers neither the similarity between states nor a semantics description of the states, where describing states using concepts and computing the similarity between them using ontologies can contribute to improve the detection of similar danger situations.
\begin{figure} [h]
\vskip 0.2in
\begin{center}
\centerline{\includegraphics [width=0.85\columnwidth]{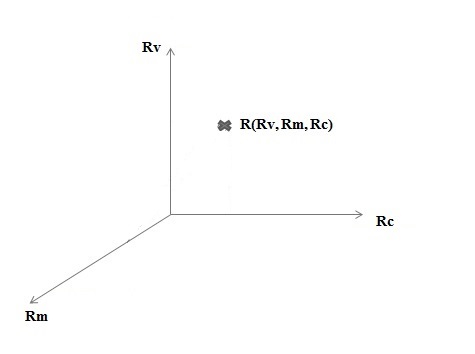}}
\caption{ Risk modelling}
\label{fig:Risk}
\end{center}
\vskip -0.2in
\end{figure}
To this end, we have aggregated three approaches for computing the risk. As it is shown in Figure~\ref{fig:Risk}, the first one is computing the risk $R_c$ using concepts. This approach permits to get the risk of the situation directly from the risk of each of its concepts. The second approach is computing the risk $R_m$ using the semantic similarity between the current situation and situations stocked in the system. $R_m$ comes from the assumption that similar situations have the same risk level. The third approach is computing the risk $R_v$ using the variance of the reward. In this case, we assume that risky situations get very low number of user's clicks.
  
In what follows, we describe the three approaches and their aggregation.

\subsubsection{Risk Computed using the Variance of the Reward}
To compute the risk of the situation using the variance of the reward, we suppose that the distribution of the Click Through Rate (the CTR is the number of clicks per recommendation) of the situations follows a normal distribution. From this assumption, and according to confidence interval theory \cite{Wald1942}, we compute the risk using Eq.~\ref{eq:Rv}. The idea here is that, more the CTR of situations is low (low number of user's clicks) more the situation is risky. 

\begin{equation}
 \label{eq:Rv}
  R_v(S^p) = \left\{ \begin{array}{rcl}
  	1- \frac{CTR(S^p) - Var}{1-Var}& \mbox{if} & CTR(S^p) > Var\\ 
       1  &   &   \mbox{Otherwise}
            \end{array}\right. 
\end{equation}

In Eq.~\ref{eq:Rv}, the risk threshold $Var$ is computed as follows :

\begin{equation}
 \label{eq:var}
Var= E(CTR(S)) - \alpha*\sigma(CTR(S)) 
\end{equation}

In Eq.~\ref{eq:var}, $\sigma$ is the variance of $CTR(S)$ and $\alpha$ is constant fixed to 2 according to Gauss theory \cite{Wald1942}.
The $CTR(S)$ is computed as follows :
\begin{equation}
 \label{eq:ctr}
CTR(S)= \frac{click(S)}{rec(S)}
\end{equation}
In Eq.~\ref{eq:ctr}, $click(S)$ gives the number of times that the user clicks in documents recommended in $S$ and $rec(S)$ gives the number of times that the system has made recommendation in the situation $S$. 

\subsubsection{Risk Computed using Concepts}
Computing the risk using concepts gives a weighted mean of the risk level of the situation concepts:
\begin{equation}
\label{eq:Rc}
R_c(S^t) = \sum_{\delta \in \Delta} \mu_{\delta} cv_{\delta}^{t} \qquad \mbox{if}\qquad CV \neq \emptyset    
\end{equation}

In Eq.~\ref{eq:Rc}, $cv_{\delta}^{t}$ is the risk level of dimension $\delta$ in $S^t$ and $\mu_{\delta}$ is the weight associated to dimension $\delta$, set out by using an arithmetic mean as follows:
\begin{equation}
 \label{eq:mu}
\mu_{\delta}=\frac{1}{|CS|}(\sum_{S^i \in CS} cv_{\delta}^{i}) 
\end{equation}
The idea in Eq.~\ref{eq:mu} is to make the mean of all the risk levels associated to concepts related to the dimension $\delta$ in $CS$. 
\subsubsection{Risk Computed using Semantic Similarity between the Current Situation and Past Situations }
The risk may also be computed using the semantic similarity between the current situation and $CS$ stocked in the system. This permits to give the risk of the situation from the assumption that a situation is risky if it is similar to a pre-defined $CS$.
 
The risk $R_m(S^t)$ obtained this way is computed using Eq.~\ref{eq:Rm} 
\begin{equation}
\label{eq:Rm}
R_m(S^t) =\left\{ \begin{array}{rcl}
		 1- B + sim(S^t,S^m)  & if \hspace{.1cm} sim(S^t,S^m) < B \\ 
              \hspace{-.5cm}  1  & otherwise 
          \end{array}\right.
\end{equation} 
In Eq.~\ref{eq:Rm}, the risk is extracted from the degree of similarity between the
current situation $S^t$ and the centroid critical situation $S^m$ (Eq. ~\ref{eq:Sm}). $B$ is the similarity threshold and it is computed using an off-line simulation. From Eq.~\ref{eq:Rm}, we see that the situation risk level $R_m(S^t)$ increases when the similarity between $S^t$ and $S^m$ increases.
The critical situation centroid is selected from $CS$ as follows:
\begin{equation}
 \label{eq:Sm}
S^m=argmax_{S^{f}\in{CS}}\frac{1}{|CS|}\sum_{S^e\in{CS}}sim(S^f,S^e)   
\end{equation}

\subsubsection{Risk Computed using the Different Risk Approaches}
The risk complete level \textit{$R(S^t)$} of the current situation is computed by aggregating the $R^c$, $R^v$ and $R^m$ as follows:
\begin{equation}
\label{eq:rst}
R(S^t)=\sum_{j \in J}\lambda_{j}R_{j}(S^t)
\end{equation}  
In Eq.~\ref{eq:rst}, $R_{j}$ is the risk metric related to dimension $j \in J$, where $J=\{m, c, v\}$; $\lambda_{j}$ is the weight associated to dimension $j$ and it is set out using an off-line evaluation.

\subsubsection{Updating the Risk Value}
\label{sec:4}
After the recommendation process and the user's feedback, the system propagates the risk to the concepts of the ontology using Eq.~\ref{eq:cv} and propagates the risk in $CS$ using Eq.~\ref{eq:CSS} :   
 \begin{equation}
 \label{eq:cv}
\forall cv \in S^t \;\; cv = \frac{1}{|CV_{cv}|}(\sum_{S^{i} \in CV_{cv}} cv_{i}^{\eta})
 \end{equation}
The idea in Eq.~\ref{eq:cv} is to make the mean of all the risk levels associated to concepts $cv$ related to situations $S^{i}$ in the user's situation historic for the dimension $\eta$. In Eq.~\ref{eq:cv}, $CV_{cv}$ gives the set of situations where $cv$ has been computed.
 \begin{equation}
 \label{eq:CSS}
R(S^t) = \frac{1}{T}(\sum_{k=1}^{k=T} R(S^{t}_{k}))
 \end{equation}
The idea in Eq.~\ref{eq:CSS} is to make the mean of all the risk levels associated to the situation $S^t$ in the user's situation historic. In Eq.~\ref{eq:CSS}, $k \in [0,T] $ gives the number of times that the risk of $S^t$ is computed.

\section{Experimental Evaluation}
\label{sec:experimental}
In order to empirically evaluate the performance of our approach, and in the absence of a standard evaluation framework, we propose an evaluation framework based on a diary set of study entries. The main objective of the experimentation is to evaluate the performance of the proposed algorithm using an off-line and on-line evaluation. In the following, we present our experimental datasets, describe how to find the optimal parameters of our algorithm and then present and discuss the obtained results. 

\subsection{Evaluation Framework}
\label{sec:evaluation} 
We have conducted a diary study with the collaboration of a software company. This company provides a history application, which records the time, the current location, the social and navigation information of its users during their application use. 

The diary study lasted 2 months and has generated 356 738 diary situation entries. Each diary situation entry represents the capture of contextual time, location and social information. For each entry, the captured data are replaced with more abstracted information using time, spatial and social ontologies  
Table ~\ref{Semantic} illustrates three examples of such transformations' results.
    
\begin{center}
\begin{table}[h]
\caption{ Diary situation}
    
\label{Semantic}       
\begin{center}
\begin{tabular}{lllll}
\hline\noalign{\smallskip}
IDS &	Users &	Time	& Place	& Client \\
\noalign{\smallskip}\hline\noalign{\smallskip}
1	& Paul	& Workday &	Paris &	Finance  client\\
2	& Fabrice &	Workday&Roubaix &	Social  client\\
3	& John    & Holiday&Paris &	Telecom  client\\
\noalign{\smallskip}\hline
\end{tabular}
\end{center}
\end{table}
\end{center}   

From the diary study, we have obtained a total of 5 518 566 entries concerning the user's navigation (number of clicks and time spent), expressed with an average of 15.47 entries per situation. Table~\ref{Diary1} illustrates examples of such diary navigation entries.
    
\begin{center}
\begin{table}[h]
\caption{Diary navigation entries}
\label{Diary1}       
\begin{center}
\begin{tabular}{lllll}
\hline\noalign{\smallskip}
IdDoc	& IDS &	Click	& Time\\
\noalign{\smallskip}\hline\noalign{\smallskip}
1	& 1	& 2	& 2' \\
2	& 1	& 4	& 3' \\
3	& 2 & 1 & 5' \\
\noalign{\smallskip}\hline
\end{tabular}
\end{center}
\end{table}
\end{center}

\subsubsection{Analysing the distribution of situations with risk levels}
To analyse the different risk levels of situations in our dataset, we have computed the risk of each situation  
using Eq.~\ref{eq:rst}. 
Then, we manually group situations depending on their risk levels in five intervals [1\%, 20\%], ]20\%, 40\%], ]40\%, 60\%], ]60\%, 80\%], ]80\%, 100\%], where 1\% corresponds to the less risky situations and 100\% corresponds to the 
most risky situations. 

\begin{figure}[h]
\vskip 0.2in
\begin{center}
\centerline{\includegraphics[width=0.70\columnwidth]{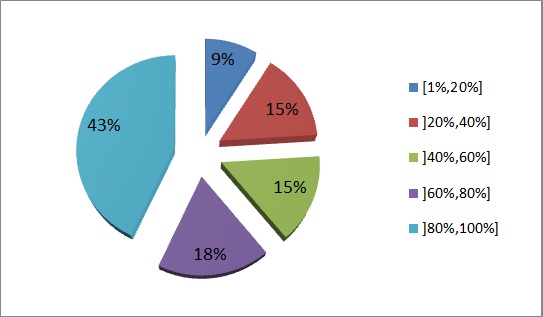}}
\caption{Distribution of situations in 5 intervals of risk levels}
\label{fig:situationcluster}
\end{center} 
\vskip -0.2in
\end{figure}

We plotted the situation distribution in 5 sectors (one for each interval) in the pie chart  
of Figure~\ref{fig:situationcluster}.  
As illustrated in the figure, we can notice that the studied domain is risky in 43\% of the situations.
To better understand the risk distribution among the situations, we further studied  
each interval according to features taken from Location, Time and Social context as well as personal information like age, gender and number of clicks.

\begin{figure}[h]
\vskip 0.2in
\begin{center}
\centerline{\includegraphics[width=1.1\columnwidth]{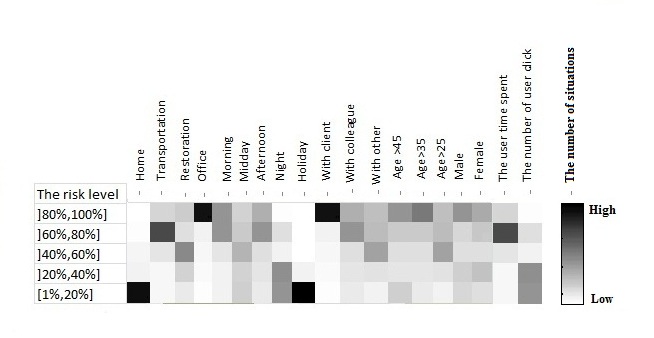}}
\caption{Information concerning situations w. r. t. their risk level}
\label{fig:distributions}
\end{center}
\vskip -0.2in
\end{figure}

The results are depicted in Fig. \ref{fig:distributions} as a 
heat map graph. Each square's gray level indicates the rate of a feature on the corresponding interval, from white (low number of situations) to black (high number of situations).

Situations of the interval ]80\%,100\%] are mostly situations where the user is in his/her office. 
The interval ]60\%, 80\%] corresponds mainly to business lunch situations; mainly home situations are in ]40\%,60\%];  
]20\%, 40\%] mostly correspond to holidays situations. We also observe that  
women have less situations with high risk level than men, and users with age between 20 and 35 are involved in few 
high risk level situations. 

One interesting finding in Fig. \ref{fig:distributions} is that the number of user's click is very low in situations with high risk level. This finding suggests some advice for content management, such as reducing recommendation in risky situations because the number of user's clicks decreases considerably.  

\subsection{Finding the Optimal B Threshold Value}
\label{sec:b}
Computing the $B$ threshold is very important because misclassifying non-$CS$ as $CS$ can be tolerated, but the opposite may be with a negative impact. 

To this end, we use a manual classification as a baseline and we compare it with the results obtained by our  
strategy.  

We first take a random sampling of 891 situations, which corresponds to 0.05\% of the  
situation entries, and manually group similar situations; then, we compare the constructed groups with the results obtained by our similarity algorithm, with different threshold values.

\begin{figure}[h]
\vskip 0.2in 
\begin{center}
\centerline{\includegraphics[width=0.9\columnwidth]{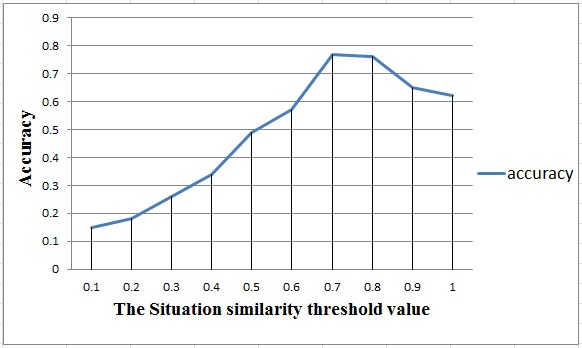}}
\caption{ Effect of $B$ threshold on the similarity accuracy}     
\label{fig:accuracy}
\end{center}
\vskip -0.2in
\end{figure}

Figure~\ref{fig:accuracy} shows the effect of varying 
$B$ in the interval [0, 1] on the overall accuracy. Results show that the best performance is obtained when $B$ has the value 0.7 achieving an accuracy of 0.769. Consequently, we use the identified optimal threshold value ($B = 0.7$) of the situation similarity measure for testing our MCARS.
\subsection{Finding the Optimal $\epsilon _{min}$ and $\epsilon _{max}$ Value}
\label{sec:epsilonmin}
In order to set out the $\epsilon _{min}$ and $\epsilon _{max}$ value, we take a sampling of 50\% of the critical 
situation entries; then we run the $\epsilon$-UCB algorithm with different $\epsilon$ values on the data.

\begin{figure}[h]
\vskip 0.2in
\begin{center}
\centerline{\includegraphics[width=0.9\columnwidth]{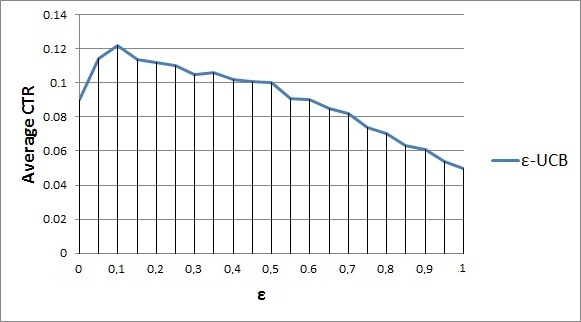}}
\caption{ Effect of $\epsilon$ value on the average CTR}
\label{fig:epsilon}
\end{center} 
\vskip -0.2in
\end{figure}

Figure~\ref{fig:epsilon} shows how the average CTR varies for $\epsilon$-UCB with the respective  
$\epsilon$. The CTR for a particular iteration is the ratio between the total
number of clicks and the total number of displays. When $\epsilon < 0.1$, there is an insufficient exploration; consequently the algorithm have failed to identify interesting documents, and have got a smaller number of clicks (average CTR).
Moreover, when $\epsilon > 0.5$, the algorithm seems to over-explore and thus loses a lot of opportunities to increase the number of clicks.

We observe from the evaluation that the best $\epsilon _{min}$ and $\epsilon _{max}$ are respectively 0.1 and 0.5. We have expected this results due to our critical situations sampling.      
\subsection{Off-line Experimental Results}
To test the R-UCB algorithm, in our experiments, we have collected, from the Nomalys' historic, a collection $Cas$ of 100000 cases $Cas^i$. The testing step consists of running the algorithm by confronting it at each iteration to a case randomly selected from $Cas$. 

For each iteration $i$ the algorithm need to select or recommend 10 documents $d \in D^i$, note that the algorithm is only confronted to the case $Cas$ where $|D|>20$ and $D \in Cas$.

We compute the average CTR (click feedback) every 1000 iterations and we have run the simulation until the number of iterations reaches 10000, which is the number of iterations where all the tested algorithms have converged. Note that, due to our goal on evaluating the RS in a periods of time or in iterative process, we have used the average CTR rather than traditional Recall used in information retrieval that do not allowed this kind of evaluation.  

\begin{figure} [h]
\vskip 0.2in
\begin{center}
\centerline{\includegraphics [width=0.95\columnwidth]{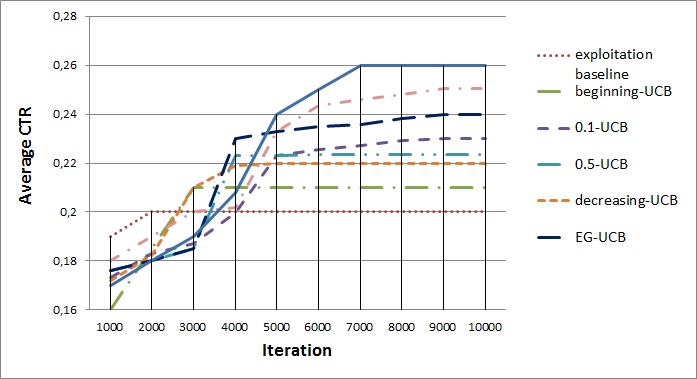}}
\caption{Average CTR for exploration-exploitation algorithms}
\label{tab:iteration}
\end{center}
\vskip -0.2in
\end{figure}

In the first experiment, in addition to a pure exploitation baseline, we have compared the R-UCB algorithm to the algorithms described in the related work : $\epsilon$-UCB and beginning-UCB, decreasing-UCB, VDBE-UCB, EG-UCB, which correspond to $\epsilon$-UCB using respectively decreasing exploration, beginning exploration, VDBE exploration and EG exploration. In Figure~\ref{tab:iteration}, the horizontal axis represents the number of iterations and the vertical axis is the performance metric.
We have parametrized the different algorithms as follows: $\epsilon$-UCB was tested with two parameter values: 0.5 and 0.1; decreasing-UCB and EG-UCB use the same set {$\epsilon_i$ = 1- 0.01 * i, i = 1,...,100}; decreasing-UCB starts using the highest value and reduces it by 0.01 every 100 iterations, until it reaches the smallest value. We have several observations regarding the different exploration-exploitation algorithms. 

For the decreasing-UCB algorithm, the converged average CTR increases as the $\epsilon$ decreases (exploitation augments).
For the 0.1-UCB and 0.5-UCB, neither a small exploration of 10\% for 0.1-UCB nor a big exploration of 50\% for 0.5-UCB give good results. This confirms that a static exploration is not interesting in this dynamic environment.
While the EG-UCB algorithm converges to a higher average CTR, its overall performance is not as good as the VDBE-UCB algorithm that considers the uncertainty of its knowledge for each situation.
 
The R-UCB and VDBE-UCB algorithms effectively have the best convergence rate, increasing the average CTR by a factor of 1.5 over the baseline for VDBE-UCB and 2 for R-UCB. This improvement comes from a dynamic exploration-exploitation trade-off, controlled by considering the situations.
 
Finally, as we expect, the R-UCB outperforms VDBE-UCB, which is explained by the good estimation of the risk based on our semantic approach. The R-UCB algorithm takes full advantage of exploration when the situations are not dangerous (non-CS), giving opportunities to establish good results when the situations are critical (CS). 

\subsubsection{Risk Level of the Situations}
To compare the algorithms in situations with different risk levels, we run the tested algorithms in the groups of situations with different risk level described in Section~\ref{sec:evaluation}. 

\begin{figure} [h]
\vskip 0.2in
\begin{center}
\centerline{\includegraphics [width=0.95\columnwidth]{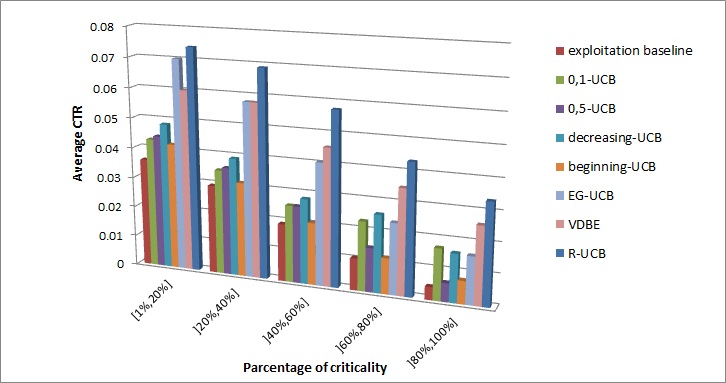}}
\caption{Average CTR of exploration-exploitation algorithms in situations with different risk levels}
\label{fig:risky}
\end{center}
\vskip -0.2in
\end{figure}

To better visualize the comparison results, Fig.~\ref{fig:risky} shows algorithms' average CTR graphs with the previous referred risk levels. Our first observation is that the R-UCB algorithm outperforms all other exploration-exploitation algorithms, at every levels. We notice that, in high risk situations, low exploration (0.1-UCB) is better than high exploration (0.5-UCB). The gap between the R-UCB results and the other algorithms increases with the risk of the situations. This improvement comes from the safety exploration made by the R-UCB.

\subsubsection{Size of Data}
To compare the algorithms when the case base is sparse in our experiments, we reduce the case base size of 50\%, 30\%, 20\%, 10\%, 5\%, and 1\%, respectively. 

\begin{figure} [h]
\vskip 0.2in
\begin{center}
\centerline{\includegraphics [width=0.95\columnwidth]{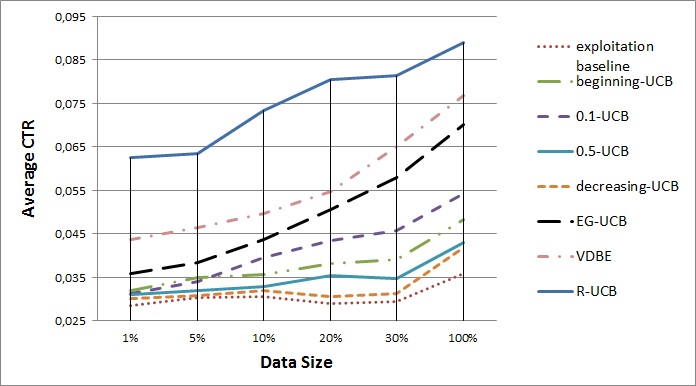}}
\caption{Average CTR for different data size }
\label{fig:datasize}
\end{center}
\vskip -0.2in
\end{figure}

To better visualize the comparison results, Fig.~\ref{fig:datasize} shows algorithms' average CTR graphs with the previous referred data sparseness levels. Our first observation is that all algorithms are useful at every level. We notice that decreasing data size does not significantly improve the performance of 0.5-UCB and 0.1-UCB. Except for exploitation baseline, beginning-UCB seems to have a rather poor performance. Its results are worse than any other strategy independently of the chosen parameters. The reason lies in the fact that this algorithm makes the exploration only at the beginning.

\subsection{On-line Evaluation Results}
Based on the results from our off-line evaluations, we have selected the most promising techniques, R-UCB, VDBE-UCB and EG-UCB, and tested them with real users in their real working environments.
We use the 3000 users of Nomalys application in full time. To qualify, participants are required to use Nomalys for more than 1 hour/week. 

To this end, we have randomly split the users in three groups. During the first week of the study, the system records the documents used by each participant without recommendation. During the second week of the study, we have equipped the first group with the RS system running the R-UCB algorithm, the second group running the VDBE-UCB, and the last group running EG-UCB algorithm. 

\subsubsection{New Document Exploration}
With a large number of users, we can not easily follow the average CTR of each user. For this reason, we use a metric to see how the usage of the RS impacted the user's usage of documents that they had not previously seen (new documents).
An increase in the usage of such documents would indicate that the system was recommending documents that were useful.

\begin{figure} [h] \vskip 0.2in 
\begin{center}
\centerline{\includegraphics [width=0.95\columnwidth]{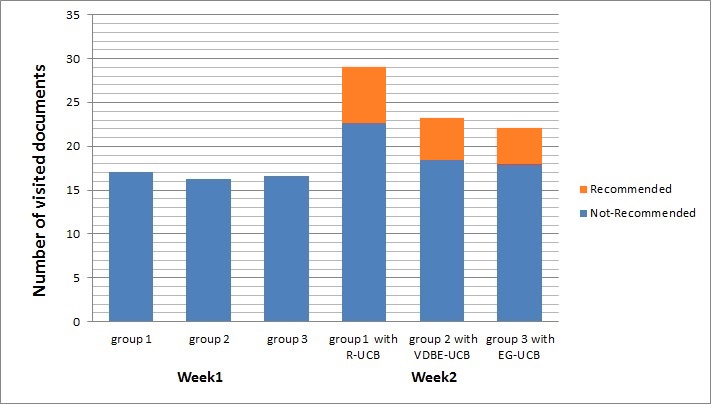}}
 \caption{Average number of new documents used in three groups without and with recommendation}
\label{fig:newdocexp} \end{center} \vskip -0.2in
 \end{figure}

By comparing the number of new documents with and without recommendation in the $1^{st}$, $2^{nd}$ and $3^{rd}$ groups, we got the impact that recommendations had on the use of new documents. Figure ~\ref{fig:newdocexp} illustrates this comparison by week and by group.
Figure~\ref{fig:newdocexp} shows a main effect for the week on the number of visited new documents. The average number of new documents used in the first week has been 17.12, 16.31 and 16.63 for groups 1, 2 and 3 respectively, and 29, 23.21 and 22.12 for groups 1, 2 and 3 respectively on the second week. Moreover, group 1 has significantly more new documents used than groups 2 and 3 in the second week. 

Based on the results illustrated in Fig. \ref{fig:newdocexp}, we can estimate the proportion of new visited documents used in the second week due to the introduction of the recommender system, and the proportion of new documents that would have been used by chance (without recommendation).
 
In the second week, with groups 1, 2 and 3 respectively, an average of 6.3, 4.7 and 4.3 documents actually appeared in the recommender list before being used (i.e. new recommended documents), which represents a very good improvement w. r. t. the first week without recommendation. 

Another interesting finding is that, excluding the average of 16.68 new documents without recommendation (first week) from the number of new documents with recommendation (second week), for groups 1, 2 and 3, and also excluding the average number of new recommended documents, we get an average number of 6.01, 1.82 and 1.13 new documents in the second week with groups 1, 2 and 3, respectively. We believe that the majority of these extra documents were discovered through the use of exr/exp in MCARS.
 
The improvement on the number of visited new documents, on one hand, corresponds most of all to recommended documents; on the other hand, an important part is discovered during recommendations, which shows an unintentional benefit of the exr/exp which promotes document discovery.

\subsubsection{R-UCB, VDBE-UCB and EG-UCB Comparison}
\begin{figure} [H] \vskip 0.2in 
\begin{center}
\centerline{\includegraphics [width=0.9\columnwidth]{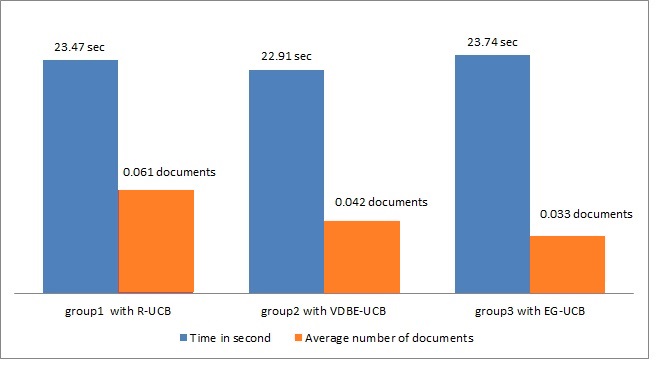}}
 \caption{Average number of recommended documents have used multiple times for each navigation session}
\label{fig:newdocument} \end{center} \vskip -0.2in
 \end{figure}

To compare the R-UCB, VDBE-UCB and EG-UCB algorithms, we look at the number of recommended documents that have been used multiple times (i. e. more than twice) in each session and the time spent in each document. Figure~\ref{fig:newdocument} shows the results.
We observe, from the figure, that the time spent in each document does not significantly change in the three groups, which means that the exploration-exploitation trade-off does not impact the user's time spent. However, it has shown that group 1 has used more new documents than groups 2 and 3, which confirms that our exploration-exploitation trade-off have allowed more exploration of documents from the users than the other strategies.   
\section{Conclusion}
In this paper, we have studied the problem of exploitation and exploration in mobile context-aware recommender systems and propose a new approach that 
adaptively balances exr/exp regarding the risk level of the situation. 

We have validated our work with a series of both off-line and on-line studies which offered promising results. Moreover, this study yields to the conclusion that considering the risk level of the situation on the exr/exp strategy significantly increases the performance of the recommender system. In considering these results, we plan to investigate  
public benchmarks. 

\bibliographystyle{abbrv} 
\small
\bibliography{sigproc} 
 
\end{document}